\documentclass[pdflatex,sn-mathphys-num]{sn-jnl}

\usepackage{graphicx}%
\usepackage{multirow}%
\usepackage{amsmath,amssymb,amsfonts}%
\usepackage{amsthm}%
\usepackage{mathrsfs}%
\usepackage[title]{appendix}%
\usepackage{xcolor}%
\usepackage{textcomp}%
\usepackage{manyfoot}%
\usepackage{booktabs}%
\usepackage{algorithm}%
\usepackage{algorithmicx}%
\usepackage{algpseudocode}%
\usepackage{listings}%

\usepackage{caption}
\usepackage{stfloats}
\usepackage{amsmath}

\usepackage{hyperref}

\usepackage{amssymb}
\usepackage{amsmath}
\usepackage{float}
\usepackage{tabularx} 
\usepackage{xcolor}

\DeclareUnicodeCharacter{2212}{-}

\begin{document}

\title[Article Title]{Quantifying Polarization: A Comparative Study of Measures and Methods}

\author*[1]{\fnm{Edoardo} \sur{Di Martino}}\email{edoardo.dimartino@uniroma1.it}
\author[2]{\fnm{Matteo} \sur{Cinelli}}\email{matteo.cinelli@uniroma1.it}
\author[1]{\fnm{Roy} \sur{Cerqueti}}\email{roy.cerqueti@uniroma1.it}
\author[2]{\fnm{Walter} \sur{Quattrociocchi}}\email{walter.quattrociocchi@uniroma1.it}

\affil*[1]{\orgdiv{Department of Social Sciences and Economics}, \orgname{Sapienza University of Rome}, \orgaddress{\street{P.le Aldo Moro, 5}, \postcode{00185}, \state{Rome}, \country{Italy}}}

\affil[2]{\orgdiv{Department of Computer Science}, \orgname{Sapienza University of Rome}, \orgaddress{\street{ Viale Regina Elena, 295}, \postcode{00161}, \state{Rome}, \country{Italy}}}

\abstract{Political polarization, a key driver of social fragmentation, has drawn increasing attention for its role in shaping online and offline discourse. Despite significant efforts, accurately measuring polarization within ideological distributions remains a challenge. This study evaluates five widely used polarization measures, testing their strengths and weaknesses with synthetic datasets and a real-world case study on YouTube discussions during the 2020 U.S. Presidential Election. Building on these findings, we present a novel adaptation of Kleinberg’s burst detection algorithm to improve mode detection in polarized distributions. By offering both a critical review and an innovative methodological tool, this work advances the analysis of ideological patterns in social media discourse.}

\maketitle
\section*{Introduction}
\label{sec:Intro}

In a world where social and political divides increasingly shape public discourse, polarization has emerged as a fundamental driver of numerous challenges. From eroding democratic debate to intensifying online conflicts, reducing social cohesion, and obstructing consensus during crises, the effects of polarization are pervasive~\cite{wef2024}. As societies grapple with these issues, understanding the mechanisms and dynamics of polarization has become a pressing concern for researchers, policymakers, and the public.

The academic literature categorizes polarization into various forms, such as affective, ideological, and issue-based polarization. Affective polarization refers to the growing emotional distance and antagonism between opposing groups~\cite{iyengar2019origins}, while ideological polarization captures the extent to which individuals' beliefs align along opposing ends of a spectrum~\cite{iyengar2012affect}. Issue polarization, on the other hand, focuses on the divergence in opinions on specific policy matters~\cite{mason2015disrespectfully}. Despite these distinctions, ``polarization" is often used more generally to describe divisions and contrasts between individuals or groups with differing viewpoints or beliefs. This broad usage highlights both the ubiquity and the complexity of the phenomenon.

While the study of polarization has expanded in recent years, much of the existing knowledge still needs to be more cohesive and disproportionately focused on the United States. Historical data from the Pew Research Center~\cite{center2014political} illustrates the deepening ideological divides in the U.S., where the proportion of individuals expressing consistently conservative or liberal opinions increased from 10\% in 1994 to 21\% in 2014, while ideological overlap between political partisans significantly declined over the same period. This growing polarization reflects a trend with global implications, as similar dynamics are observed in other democracies, albeit with local variations~\cite{falkenberg2024patterns}.

The role of social media in political polarization has become a subject of intense debate, characterized by opposing views that, ironically, reflect polarization within the academic and public discourse itself. Some researchers argue that social media platforms amplify divisions by fragmentation~\cite{falkenberg2022growing, sunstein2004democracy, wsj_facebook_files_2021}, echo chambers~\cite{cinelli2021echo} and, eventually, echo platforms~\cite{cinelli2022conspiracy, di2024characterizing}. Others contend that these platforms provide opportunities for democratic engagement and cross-partisan dialogue~\cite{barbera2015tweeting}, fostering interactions that might not occur in offline settings~\cite{north2019social}. The divergence in these perspectives underscores the need for robust methodologies to analyze and measure polarization effectively in online environments.

Despite its importance, measuring polarization remains a challenging task. Bramson \emph{et al.}~\cite{bramson2017understanding}, in a review of meanings and methods associated with polarization, identified nine different basic ``types" of polarization that one can consider when studying measures of beliefs distributed on a normalized spectrum along one dimension (simply put, as a histogram representing the number of individuals holding a specific belief), pointing out how ``the most common measure of polarization in the political literature is probably bimodality". Following this paradigm, a variety of studies have used statistical techniques to compute the extent of deviation from unimodality in a sample as a proxy to quantify political polarization \cite{falkenberg2022growing, de2022modelling, kim2022democracy, pavlopoulos2023distance}.

However, while the concept of bimodality is theoretically appealing, its practical implementation through algorithmic methods could be more straightforward. The challenge becomes particularly acute when analyzing large datasets, where manual inspection of individual distributions is infeasible, necessitating the development of efficient and reliable measures.

The complexity of measuring polarization is further compounded by the diverse contexts in which it manifests. Political discussions on social media, for example, are shaped by factors such as platform algorithms, user behavior, and the structure of online networks \cite{cinelli2021echo, hussein2020measuring, avalle2024persistent}. This multifaceted nature of polarization demands tools to account for the underlying distributional properties of ideological beliefs and the specific characteristics of the digital environments in which these beliefs are expressed.

In this paper, we aim to address these challenges by comprehensively reviewing existing measures for quantifying polarization. We critically examine their theoretical underpinnings, strengths, and limitations, focusing on their applicability to large-scale datasets. To illustrate the practical implications of these measures, we conduct a systematic evaluation using synthetic data and apply them to a case study on political discussions on YouTube during the 2020 U.S. Presidential Election. This empirical analysis highlights the varying performance of the measures and underscores the need for more nuanced tools to capture the complexities of polarization in online discourse.
To address the limitations of existing approaches, we introduce an adaptation of Kleinberg’s burst detection algorithm, originally designed for time-series analysis, as a novel method for mode detection in polarized distributions. By integrating this approach with traditional measures, we enhance their interpretability and provide a more comprehensive framework for analyzing ideological patterns in social media discussions.

Through this study, we contribute to the growing body of research on polarization by offering a critical synthesis of existing methodologies and an innovative tool for future investigations. Our findings hold implications for researchers seeking to understand the dynamics of polarization and for policymakers and platform designers aiming to mitigate its adverse effects on democratic processes and social cohesion.

\section*{Description of the selected measures}

Trying to capture political polarization by quantifying the extent of departure from unimodality of a distribution of political beliefs, as we will see later, we set on to review five different measures that deal with such applications. While an in-depth theoretical evaluation of the different measures goes beyond the scope of this study, we will provide the necessary background to understand the working mechanisms of the five measures we tested and employed.

\subsection*{Bimodality Coefficient}
The Bimodality Coefficient's derivation is attributed to Warren Sarle, and was first presented in 1990 in the User Guide for the statistical software SAS~\cite{sas1990sas}. The asymptotic version of the coefficient, however, appears in a botanical study published in 1987 by Ellison \emph{et al.}~\cite{ellison1987effect}, citing as a source the 1982 version of the SAS User Guide, now largely unobtainable.

Historical accuracy aside, the Bimodality Coefficient (which we will refer to as BC) is computed as:

\begin{equation}
    BC = \frac{\gamma^2 + 1}{\kappa}
\end{equation}

Where $\gamma$ and $\kappa$ are the skewness and kurtosis of a population, here defined as the standardized fourth moment around the mean. Intuitively, we can conceptualize the skewness as as the degree of asimmetry observed in the distribution, while the kurtosis indicates how ``heavy" are the tails of the distribution.
The value of the coefficient will therein lie between 0 and 1: the logic behind it is the fact that, if a bimodal distribution has very low kurtosis or an asymmetric character, the coefficient will increase.

When dealing with a sample distribution $b$ having $n$ observations, the BC retains its ease of implementation, as it only requires the sample size $n$, the sample's skewness, and the sample's excess kurtosis to be computed:
        \begin{equation}
        BC(b) = \frac{g^2+1}{k+3 \frac{(n-1)^2}{(n-2)(n-3)}}      
        \end{equation}
where $g$ refers to the sample's skewness and $k$ to the sample's excess kurtosis, with both moments being corrected for sample bias \cite{pfister2013good}. The value obtained from this coefficient on a given sample can be compared to an empirical benchmark value of ${BC_{bench} \approx 0.55}$, which is the value of the coefficient that would be expected from a uniform distribution. Higher values suggest a bimodal or multimodal distribution \cite{knapp2007bimodality}, while a lower one suggests a unimodal one.

\subsection*{Hartigan's Dip Test}
Hartigan's dip test, as originally proposed in 1985~\cite{hartigan1985dip}, was meant to test the null hypothesis of unimodality against the alternative hypothesis of multimodality (even though it can be utilized in the bimodal context), with the null of an asymptotic uniform distribution. It does so by computing how different the empirical cumulative distribution function of the observed data is with respect to a unimodal one, which is considered as a density function with a single inflection point between concave and convex segments.

More in detail, for a sample distribution with given sample size $n$, the Hartigan's dip statistic (HDS) is computed by considering the maximum difference between the observed distribution of data and a uniform distribution with the same sample size $n$, chosen to minimize this maximum difference. Following this reasoning, by repeatedly sampling from the uniform distribution, a sampling distribution over these differences is produced; thus, a multimodal distribution is one in which the Hartigan's Dip Statistic is equal or greater than the 95th percentile between all the sampled values, meaning that a multimodal distribution has statistically significant disparities in its ECDF~\cite{freeman2013assessing}.

Common implementations of the dip test in statistical software produce the \textit{dip statistic} and its associated p-value. The dip test statistic increases as the distribution departs from a unimodal distribution, while for the p-value, ``values less than .05 indicate significant bimodality, while values higher than .05 but less than .10 suggest bimodality with marginal significance"~\cite{freeman2013assessing}.

\subsection*{Distance from Unimodality (DFU)}
This recent measure, introduced in 2023 by Pavlopolous and Likas~\cite{pavlopoulos2023distance}, was specifically built to compute political polarization in sample data. To do so, it requires to bin the data into $K$ ordinal categories, and compute the extent of deviation from the ``unimodality rule" (as defined in what follows) on the resulting histogram. Suppose a data set $X = \{x_1, ..., x_n\}$, where each $x_i$ can take one of $K$ ordinal ratings, i.e., $x_i \in \{O^1, ..., O^K\}$, and let $f = (f_1,...,f_K)$ be the relative frequencies of the K ratings of X. Now, the discrete distribution $f$ is unimodal if it has a single mode: as such, there should be a maximum value $f_m$, while the values $f_i$ monotonically decrease as one moves away from the mode $m$. Based on this, the difference values $d = (d_2,...,d_K)$ are computed as follows:
        \begin{equation}
            d_i =
            \begin{cases}
            f_i - f_{i-1} & \text{$m<i<K$}\\
            f_i - f_{i+1} & \text{$2<i<m$}\\
            0 & \text{$i=m$} \,.
            \end{cases}       
            \end{equation}
        Then, the DFU value is defined as the maximum $d_i$ value:
        \begin{equation}
            DFU = \max{d} \,.
        \end{equation}
        In the case of a unimodal histogram, $DFU = 0$ since all $d_i$ will be negative, except for $d_m$, which is gonna be equal to 0. On the other hand, $d_i > 0$ indicates a deviation from unimodality. The maximum value the DFU can take is 0.5, achieved when the histogram is made up of two bins having the same height (i.e. frequency). Note that for uniform histograms, $f_{i}=\frac{1}{K}, \forall i$, hence $d_i = 0$ and $DFU = 0$, meaning that uniform distributions will be considered as unimodal.

\subsection*{Van der Eijk's Coefficient of Agreement}

Originally proposed in 2001 by Cees Van der Eijk~\cite{van2001measuring}, the coefficient of agreement ($A$) operates on ordered rating scales (or, more simply, on frequency vectors) and yields a value lying in the interval $[-1, +1]$. A coefficient score of $0$ is exclusively observed in cases of uniform distributions, while scores below $0$ are seen when the distribution approximates a bimodal one, with a score of $-1$ indicating perfect disagreement, i.e., when all of the values are equally distributed in the two furthest bins. Conversely, scores above $0$ suggest unimodality, where $+1$ signifies perfect agreement, which is indicative of a concentration of the values within a single bin on the rating scale. The coefficient, for a ordered rating scale with $K$ categories derived from a given sample distribution, is computed as:

\begin{equation}
            A = U \cdot (1- \frac{(S-1)}{(K-1)}) \,,
\end{equation}

where $S$ and $K$ are, respectively, the number of non-empty categories in the rating scale, and the total number of categories in the rating scale. The parameter $U$ represents a measure of unimodality, computed as:

\begin{equation}
            U = \frac{(K-2) \cdot TU - (K-1) \cdot TDU}{(K-2) \cdot (TU + TDU)} \,,
\end{equation}

where $TU$ refers to the number of triples of categories conforming to unimodality, and $TDU$ is the number of triples of categories deviating from unimodality.

The measure iteratively breaks down the original rating scale into semi-uniform layers, and computes a score $A_i$ for each layer $i$. The final score is thus computed as 
\begin{equation}
A_{overall} = \sum_{i=1}^{n} w_i \cdot A_i \,,
\end{equation}
where $w_i$ is the weight associated with the layer $i$.

As the scale of the coefficient is not intuitive to interpret when expressing polarization, we transform the scores as $\frac{(1-A)}{2}$. The midpoint $0.5$ thus expresses a uniform distribution, while scores higher (or lower) than it indicate a bimodal (or unimodal) distribution, corresponding to a polarized (or unpolarized) situation.

\subsection*{Balance}
Developed to supplement the previously mentioned Bimodality Coefficient~\cite{de2022modelling}, this intuitive measure divides an array of leaning scores in two sets, one containing values below zero, and the other containing values higher than zero. The balance score $\beta$ is then computed, for a given array, as follows:
\begin{equation}
    \textit{Balance} = \frac{min(c_1, c_2)}{max(c_1, c_2)} \,,
\end{equation}
where $c_1$ and $c_2$ represent the number of samples in the two sets. The resulting value of \textit{Balance} will then be comprised in the interval $[0,1]$, with $1$ suggesting perfect ``balance", meaning an equal number of elements between the two sets.

\begin{figure}[h]
  \centering
\includegraphics[width=.8\textwidth,height=.8\textheight,keepaspectratio]{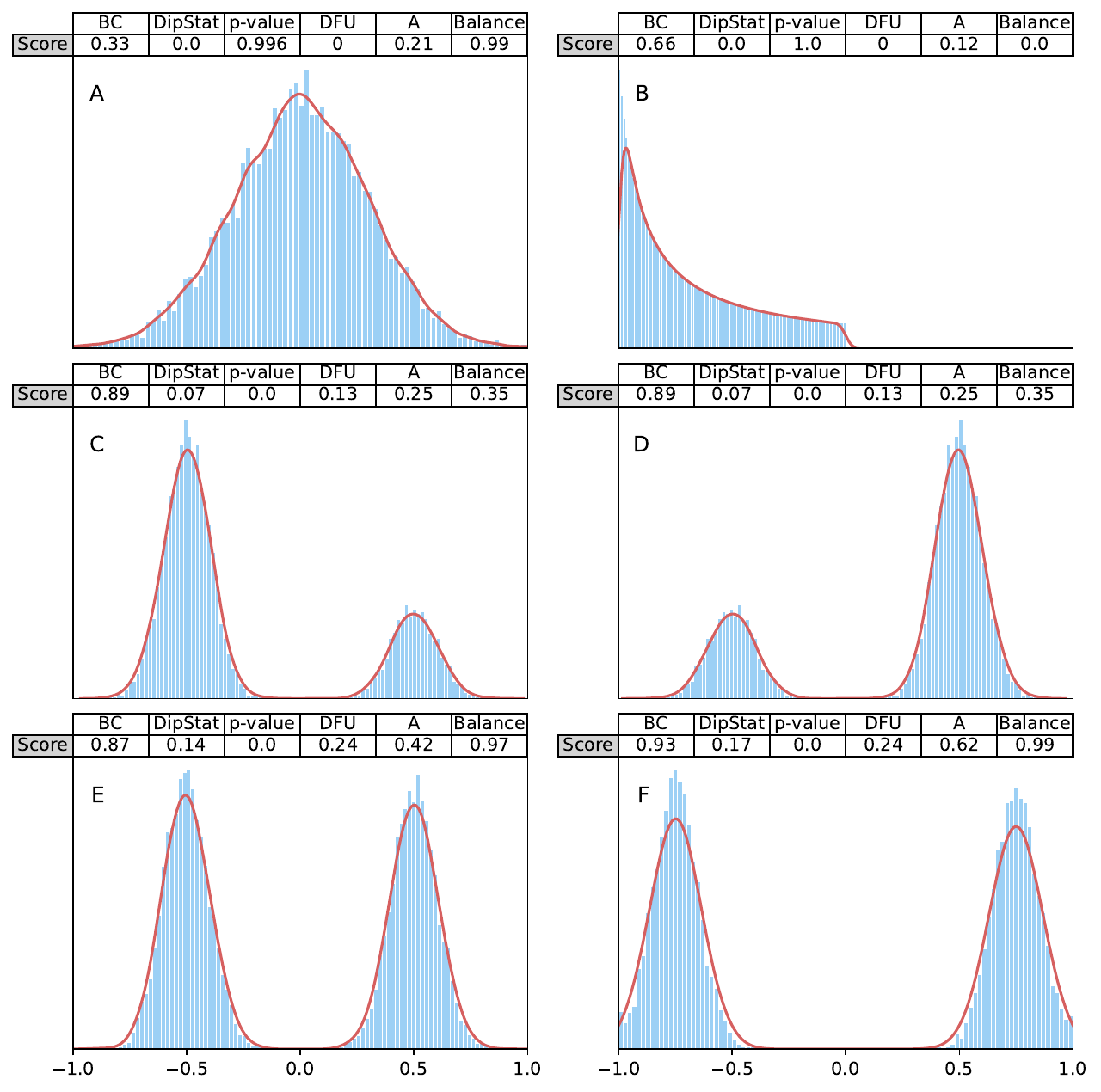}  \caption{Measures' performance on synthetically generated distributions. The tables above each panel show the score for the different measures. All of the distributions are in the interval $[-1, +1]$. An estimation of the density has been added on top of the histograms for ease of visualization.}
  \label{fig:polarcomp}
\end{figure}

\section*{Comparison on synthetic data}
To illustrate the behavior of the selected measures in a controlled setting, we compare their performance on synthetic datasets, each comprising $n=10000$ samples. The distributions are displayed in Figure~\ref{fig:polarcomp}; note that the sample in Figure~\ref{fig:polarcomp}, Panel A comes from a Gaussian sample having parameters $\mu = 0, \sigma = 0.3$, the one in Panel B represents a monotonically decreasing pattern generated using exponential decay, while Panels C, D, E and F contain distributions generated from combining random samples coming from two Gaussian distributions. The details regarding the generation of the latter 5 samples can be referred to in the Methods section. For the purpose of this study, the two measures necessitating the transformation of sample data into frequency vectors (namely, Van der Eijk's A and the DFU) are computed using eight bins. 
In all cases, the synthetic samples are truncated to be strictly comprised in the interval $[-1, +1]$. The reason for this truncation is two-fold: first, to ensure a correct functioning of the \textit{Balance} measure, which necessitates symmetrical upper and lower bounds around zero; second, to maintain consistency with the empirical analysis, where distributions are also constrained to this range.

The results across different distributions, shown in Figure~\ref{fig:polarcomp}, reveal notable characteristics. First, the Bimodality Coefficient struggles with skewed unimodal distributions due to its reliance on skewness in the numerator. This dependency reduces its robustness compared to other measures. A particularly severe failure occurs with monotonically decreasing distributions that are skewed yet unimodal (Panel B). Despite being unimodal by definition, the Bimodality Coefficient often classifies such distributions as strongly bimodal. This limitation warrants caution, as skewed distributions are common in real-world data, particularly in scenarios involving opinion polarization in ``bipartisan" systems with imbalanced populations (e.g., the pro-vax/anti-vax debate)~\cite{schmidt2018polarization}. Misclassification in these cases can distort the perceived degree of polarization.

Another observation is that Van der Eijk's $A$ is highly sensitive to the distance between modes. For instance, a distribution that is visually bimodal (Panel E) may not be identified as such by $A$, whereas a similar distribution with more widely separated modes (Panel F) crosses the $A>0.5$ threshold to be classified as bimodal. This contrasts with the DFU, which focuses solely on deviations from the ``unimodality rule" and is unaffected by the spatial separation of modes. Depending on the application, this sensitivity of $A$ to mode distance could either be advantageous or undesirable, depending on whether polarization is defined in terms of both group separation and modality.

Finally, all measures correctly identify the distribution in Panel A as unimodal. When applied to mirrored distributions (Panels C and D), all measures behave as expected, producing identical scores for both cases.

\section*{Case study on political conversations}
With the aim of contributing to the ongoing debate regarding the prevalence of polarization in online political discourse, we test the aforementioned measures using as a case study political discussions on YouTube, a platform widely used by internet users to get news and opinions~\cite{stocking2020many}, and that has drawn significant attention in studies focused on online social dynamics~\cite{wu2021cross, ribeiro2020auditing, hammas2023tuberaider, papadamou2021over}. By employing web scraping techniques, we collected 76 million comments made by almost 9 million unique users, which were extracted from $159,000$ political YouTube videos uploaded by $988$ US-based news channels, the list of which was collected between January 2020 and August 2020, the period of the latest US Presidential Election campaign, by Wu \& Resnick~\cite{wu2021cross}. The time frame was chosen for its relevance and extensive media coverage of the events leading up to the election day, and deemed favorable in fostering a rich environment for political discussions and debates. Using the classification provided in~\cite{wu2021cross}, each channel was labeled as having either a left, center, or right-leaning political bias. Table~\ref{tab:datasetbreakdown} provides a breakdown of the collected data. More details regarding the data collection process and the filtering of relevant videos are available in the Methods section.

\begin{table}[h!]
    \centering
    \renewcommand{\arraystretch}{1.5} 
    \caption{Breakdown of the data set showing how channels, videos, and comments are distributed over sources of different leaning.}
    \begin{tabular}{|l|c|c|c|c|}
        \hline
        \multicolumn{5}{|c|}{\textbf{Data set breakdown}} \\
        \hline
        \textbf{} & \textbf{Left-leaning} & \textbf{Center-leaning} & \textbf{Right-leaning} & \textbf{Total}\\
        \hline
        Channels   & 417 & 163 & 408 & 988\\
        \hline
        Videos & 84k & 21k & 54k & 159k\\
        \hline
        Comments & 42M & 7M & 27M & 76M\\
        \hline
    \end{tabular}
    \label{tab:datasetbreakdown}
\end{table}

Labeling users to infer their political leaning is a critical step and matter of active research encompassing a variety of possible approaches~\cite{barbera2015birds, cossard2020falling, zhou2011classifying, hemphill2020two, preoctiuc2017beyond, wu2021cross, falkenberg2022growing}.
To infer users leaning we track their commenting activity across the whole data set. This solution stems from the assumption that users who are often seen commenting under right-wing channels are likely to be right-leaning, and vice versa, an assumption that, despite clear limitations, finds roots in psychological theories such as selective exposure~\cite{stroud2010polarization}. We assign a score of either $\{-1, 0, +1\}$ to each comment, depending on whether it was posted under a video uploaded by a left, center, or right-leaning channel, respectively. For a user $i$  who posts $n$ comments $C_i = \{c_1, c_2,...,c_n\}$, each comment $c_j$ is associated with one of these numeric values. The political leaning $x_i$ of user $i$ is then defined as the average of the leanings of all their comments: $x_i \equiv \frac{\sum^{n}_{j=1}c_j}{n}$. This returns a leaning score in the interval $[-1, 1]$, where a value of $-1$ ($+1$) indicate that the user solely posted comments under left-leaning (right-leaning) videos.

\begin{figure}[H]
  \centering
\includegraphics[width=\textwidth,height=\textheight,keepaspectratio]{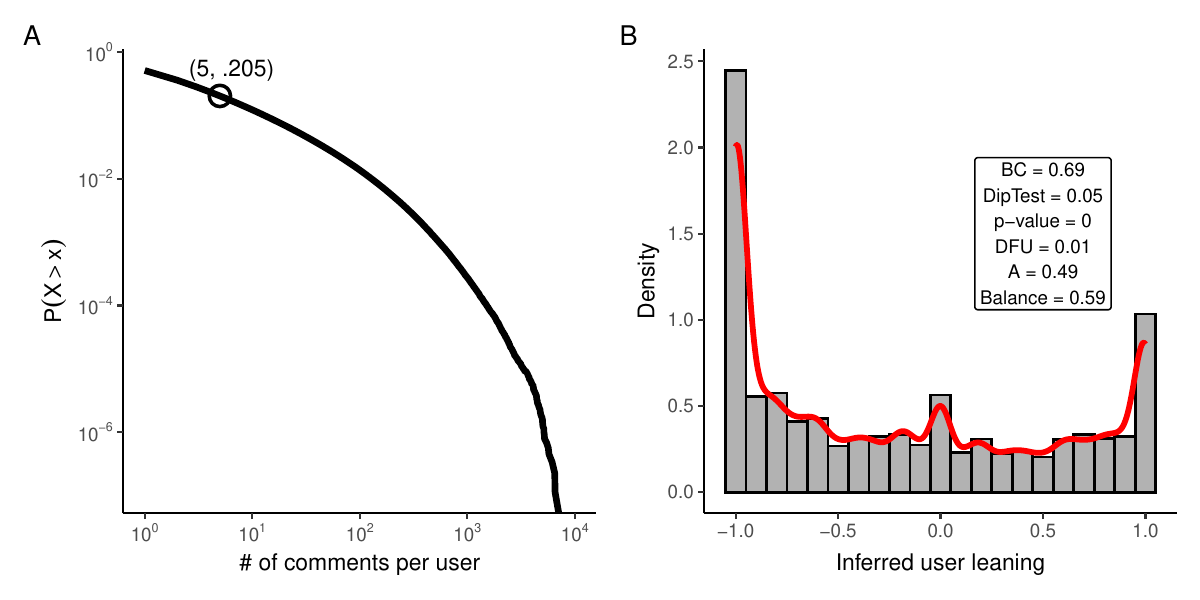}
 \caption{Empirical Complementary Cumulative Distribution Function (ECCDF) of the number of comments per user and density of the inferred users' leaning. Most of the users in the data set post few comments, while only a minority are relatively active: the median number of comments per user is $2$ (mean = $8.7$), and $48\%$ of the user base in the data set only posted one comment. The users' leaning instead is skewed towards left-leaning values. The scores of the measures computed on the inferred user leanings paint a polarized scenario.}
\label{fig:comments_and_leaning}
\end{figure}

In this regard, it is important to notice how we only retain ``active" users in the analysis, defined as those exhibiting sustained engagement (i.e. at least five comments over the whole time window). As shown in Panel A of Figure~\ref{fig:comments_and_leaning}, while about 8.8 million unique users are present in the dataset, only a fraction of them posted more than five comments, leading to the identification of $2.1$ million active users.
Referring to the distribution of users' leaning in Panel B of the same figure, it is clear how the sample is skewed towards left-leaning values, with a mean leaning score of $-0.205$, and with $59.6\%$ of users having a leaning score lower than 0. The five measures we consider indicate a certain level of polarization among the user population. While this can be observed visually when looking at a single distribution, the purpose of using these five measures goes beyond merely identifying whether the overall population is polarized. By focusing on individual comment sections, which we are gonna utilize as a proxy for a ``political discussion", we can study the degree of polarization within each one, which would be impractical - if not downright unfeasible - to do manually: if we detect a low degree of polarization among the comment sections, it may indicate segregation between users, suggesting that there is little to no political cross-talk between individuals having different political views: an urgent concern, as this lack of interaction can potentially foster toxicity in online spaces~\cite{avalle2024persistent}.

\subsection*{Applying the measures}

We evaluate the five selected polarization measures by applying them to the distributions of user leanings derived from the comment sections of videos with at least five comments from unique active users, totaling $117,457$ comment sections. For each video, we generate a vector of leaning scores, where each score corresponds to the leaning of a unique user commenting on the video. The five measures are then computed based on these vectors.

\begin{figure}[H]
  \centering
\includegraphics[width=\textwidth,height=.7\textheight,keepaspectratio]{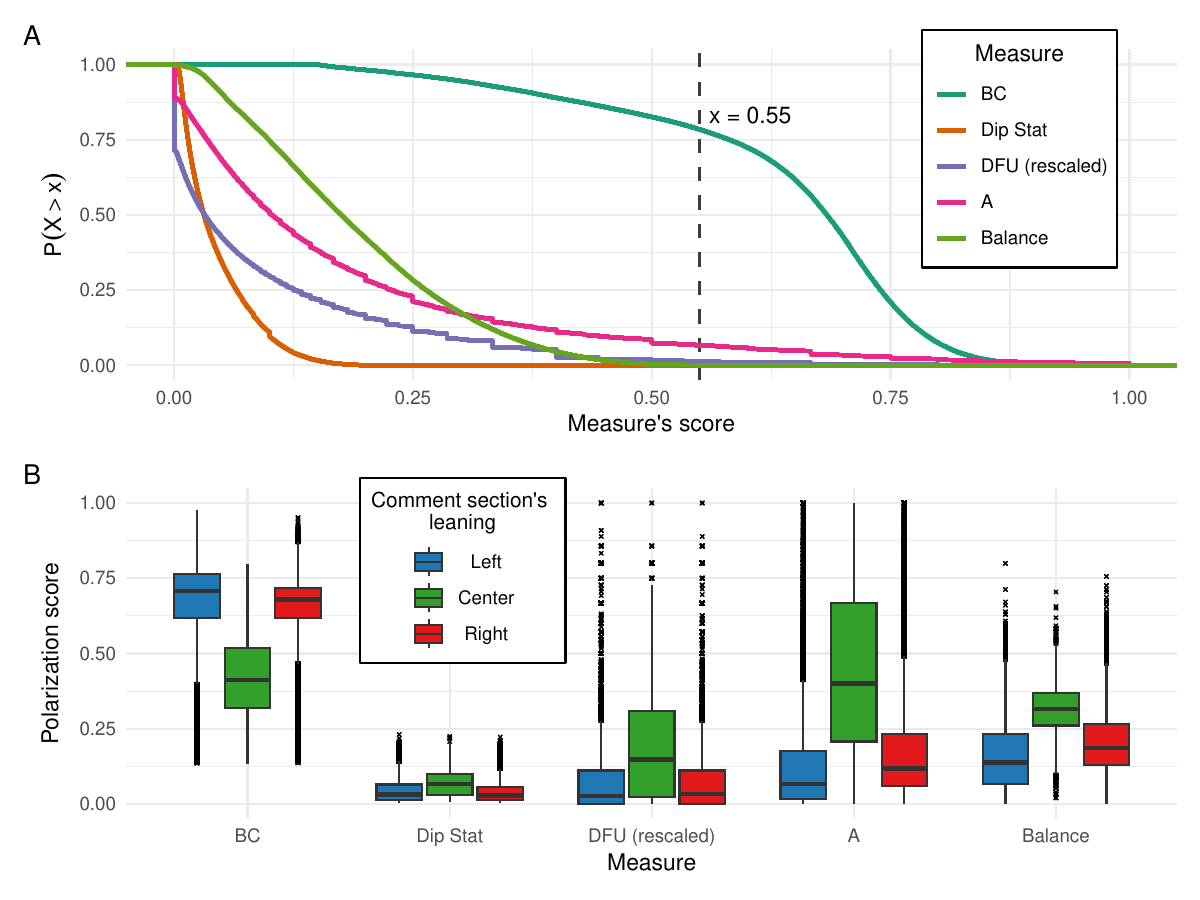}
  \caption{ECCDF of the polarization measures' scores over the leaning array of the comment sections (Panel A). The vertical line at $x=0.55$ indicates the benchmark threshold for a bimodal distribution when considering the Bimodality Coefficient.\\
  The boxplot of the distribution of the different measures differentiated by leaning (Panel B) shows us how all of the measures - bar the Bimodality Coefficient - consider center-leaning comment sections as more polarized than their left and right-leaning counterparts.\\
  The DFU scores in both panels are rescaled to the interval $[0, 1]$ to ensure visual consistency with the other measures.}
  \label{fig:casestudy}
\end{figure}

As shown in Figure~\ref{fig:casestudy}, the results vary depending on the chosen measure. Note how the scores for the DFU in the figure have been rescaled to the interval $[0, 1]$ (from the original $[0, 0.5]$) to ensure visual consistency with the other measures. 

A notable proportion of videos ($78.42\%$) exhibit a Bimodality Coefficient score exceeding $0.55$ (mean~=~$0.64$, median~=~$0.68$), generally suggesting bimodality and, by extension, polarization. However, this finding is inconsistent with the results of other measures and may reflect the BC's previously discussed vulnerability to skewed distributions.

In contrast, the Dip Statistic and DFU largely classify the comment sections as either unpolarized or weakly polarized. The Dip Statistic shows a mean value of $0.044$ and a median of $0.031$, while the DFU yields a mean value of $0.05$ and a median of $0.02$, with $28.5\%$ of comment sections having DFU~=~$0$. Expanding on the Dip Statistic, only $5304$ comment sections ($4.52\%$ of the total) exhibit a p-value associated with Hartigan's Dip Statistic below $0.1$, and $3479$ ($2.96\%$ of the total) have a p-value below $0.05$. These results reinforce the observation that the majority of comment sections are not polarized.

The other two measures present additional perspectives. Van der Eijk's $A$ has a mean value of $0.18$ (median = $0.15$), with only $1.04\%$ of comment sections exceeding the polarization threshold of $A > 0.5$. In contrast, the \textit{Balance} measure has a mean value of $0.17$ (median = $0.102$), with $11.35\%$ of comment sections scoring exactly $0$.

Interestingly, all measures except the Bimodality Coefficient indicate that videos with center-leaning audiences tend to be more polarized than those with left- or right-leaning audiences. This suggests that centrist videos may foster greater political cross-talk, serving as a potential meeting ground for divergent perspectives.

The results, however, remain fragmented, raising the question of which measure is most appropriate for assessing polarization. The Bimodality Coefficient, while intuitive and computationally efficient, struggles with skewed distributions common in such data. \textit{Balance}, on the other hand, may oversimplify the concept of polarization, potentially obscuring important dynamics, and appears better suited as a complementary tool to enhance the interpretation of other measures. The Dip Test is grounded in strong theoretical principles, but its Dip Statistic alone provides limited insight and ideally requires the accompanying p-value for meaningful interpretation.

Both the DFU and $A$ rely on binning user leanings into predefined intervals, introducing flexibility but also necessitating careful parameter selection based on the specific definition of polarization. It is worth noting, however, that $A$ is sensitive to the distance between distribution modes, which could lead to unexpected results in cases where the modes represent ideologically proximate groups. Researchers must account for this sensitivity when using $A$ to avoid mischaracterizing polarization levels.

\clearpage
\section*{Supplementing the measures: Burst detection}

Using both synthetic and social media data we have so far shown how sensitive the results are when we apply different measures designed to quantify political polarization within a dataset. In the case of YouTube data, while all but one of the measures tell us that, on average, our comment sections are not polarized, there is a certain degree of variance (due to disagreement among different measures) in the results. Thus, choosing the appropriate measure or performing a thorough comparison is far from being trivial, especially in absence of a ground truth.

Measures such as Hartigan's Dip Test, the DFU, and the Bimodality Coefficient are well-suited for assessing whether a distribution departs from unimodality but provide limited information about the overall ``shape" of the distribution. For instance, with the DFU, we know that a distribution is perfectly bimodal only when the score reaches its upper bound of 0.5. Van der Eijk's $A$, on the other hand, offers more detailed insights into the distribution's shape and is generally easier to interpret than the DFU or the Bimodality Coefficient. In contrast, \textit{Balance} is less informative about deviations from unimodality, as its primary focus is on directly quantifying polarization rather than analyzing the shape of the distribution.

Drawing from classical statistical examples, such as Anscombe's quartet~\cite{anscombe1973graphs}, a useful approach to understanding polarization would be to visualize the data and count the modes in the distribution. However, this method becomes impractical when dealing with large datasets, such as the previously introduced YouTube comment sections. A more scalable solution would involve algorithmically identifying the number of modes in the data, thereby eliminating the need for manual inspection while ensuring a principled approach to mode detection.

\begin{figure}[h]
  \centering
\includegraphics[width=\textwidth,height=.3\textheight,keepaspectratio]{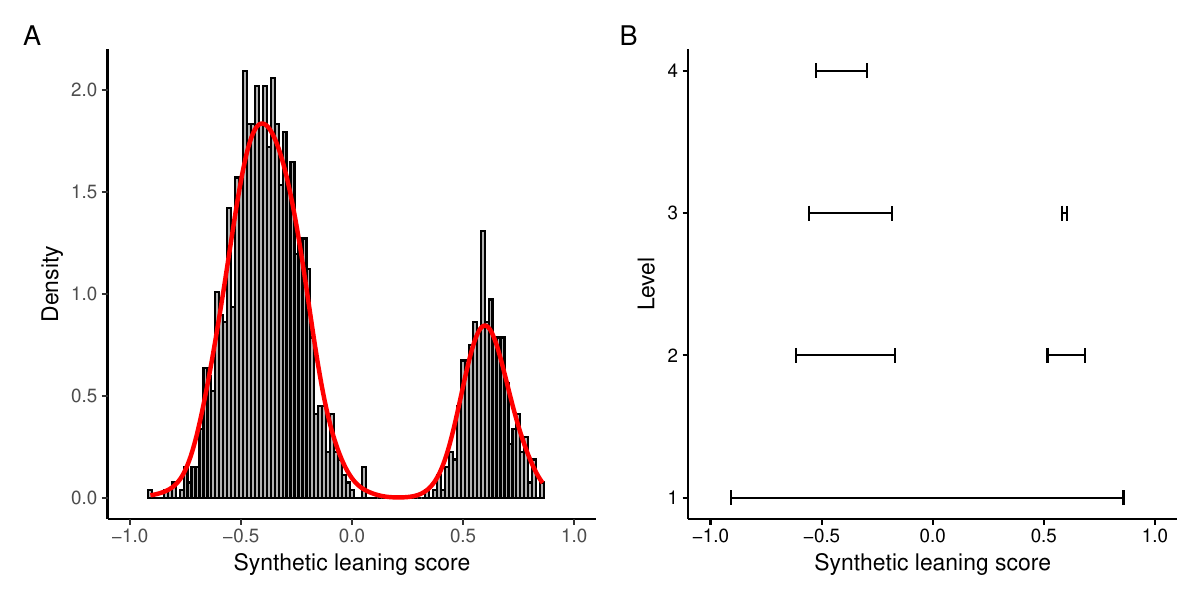}
  \caption{Results of Kleinberg's burst detection algorithm, toy example. Panel A refers to the toy distribution, while Panel B illustrates the burst detection output. The algorithm detects different peaks of varying intensity coinciding with the distributions' modes. The level of intensity 1 always spans the entire sample range.}
  \label{fig:kleinberg_toy}
\end{figure}

For this reason, we start by adapting a method initially conceived for burst detection in time series and we use it for counting the modes in a distribution. The basic idea behind the re-purposing and extension of this method is that a density profile of a time distribution can be interpreted as a high frequency time series where bursts (i.e. peaks) are the relevant modes to be identified. To do so, we employ Jon Kleinberg's burst detection algorithm~\cite{kleinberg2002bursty}, an algorithm originally intended to work with series of discrete events with known timestamps, which utilizes an infinite Markov model to detect peaks in the events activity. Two key parameters in the algorithm are $s$, controlling the amount of smoothing applied to the event data, and $\gamma$, a coefficient that determines the level of deviation from a baseline that must be exceeded for an event to be classified as a burst. Here, we define each of our comments in a comment sections as an ``event", and utilize the leaning score of the user who posted the comment as a ``timestamp". Due to the way that the inter-arrival time is computed in the Markov model, the presence of two events with the exact same timestamp is not allowed: while this issue is fairly common in many of the comment sections considered, we solve it by simply rounding the leaning scores to the third decimal, and iteratively adding a small constant $ \epsilon = 0.0001$ to the non-unique scores, to ensure the correct functioning of the algorithm without a significant distortion of the data. This procedure produce a distribution of leanings ordered on the x-axis as a time series. 
The burst detection algorithm applied on the data return an intensity level in the range $[1,+\infty)$ for each portion of the density profile which can be interpreted as the burstiness of the portion of data taken into account, as illustrated in Fig.~\ref{fig:kleinberg_toy}.

To account for the fact that peaks which are close together might reasonably be intended as a single peak, and not as two separate ones, we complement the explained method with the following procedure: after applying the burst detection algorithm and identifying bursts at various levels, we compute the Highest Density Interval (HDI) of the leanings' distribution with a specified coverage probability $\alpha$. The length of this interval is then multiplied by a constant $k$ (where $k \in (0,1)$) to obtain the aggregation parameter $\phi$. Next, for any two bursts of the same level, if the distance between the endpoint of one burst and the starting point of the subsequent burst is less than $\phi$, the two bursts are aggregated into a single burst. More formally, let $B_1 = [s_1, e_1]$ and $B_2 = [s_2, e_2]$ be two consecutive bursts of the same level, where $s_i$ and $e_i$ represent the start and end points of burst $i$. The bursts are aggregated according to the following rule:
$$ B_{\text{agg}} = \begin{cases}\{[s_1, e_2]\}, & \text{if } s_2 -e_1 < \phi \\\{B_1, B_2\}, & \text{otherwise}\end{cases} $$ After this aggregation step, if a burst at a higher level is entirely contained within a burst at a lower level, the lower-level burst is removed from the results. This prevents the same peak from being counted multiple times at different levels. 
After completing the aggregation step, we define a peak, following previous studies~\cite{avalle2024persistent}, as any burst with an intensity level equal to or greater than 3.\\
\newline

\begin{figure}[t]
  \centering
\includegraphics[width=\textwidth,height=\textheight,keepaspectratio]{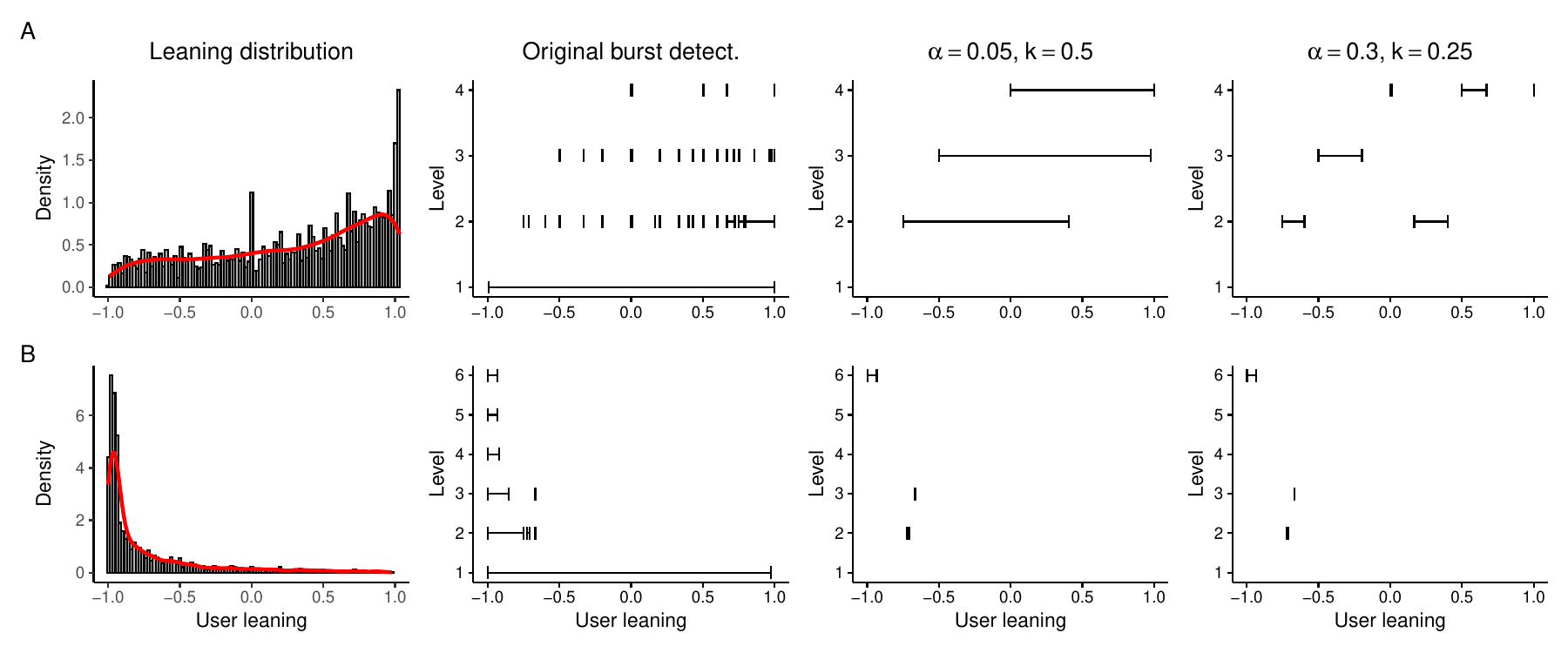}
  \caption{Results of Kleinberg's burst detection algorithm, and the subsequent burst-aggregating procedure, on different leanings' distributions. The burst-aggregating procedure are presented for two different levels of $\alpha$ and $k$. In Panel A, the different values of $\alpha$ and $k$ produce substantially different outputs (identifying, respectively, two and four bursts of intensity higher than 3), while in Panel B there is total agreement regarding the burst aggregating procedure.}
  \label{fig:kleinberg_real_distros}
\end{figure}

The results of this procedure associated to selected examples are shown in Figure~\ref{fig:kleinberg_real_distros}. In these examples, the leaning distributions are first analyzed using Kleinberg’s burst detection algorithm, followed by further refinement based on the amplitude of the distribution's HDI at two different levels. Generally, as distributions become noisier, lower values of $\alpha$ in the HDI or $k$ in the computation of $\phi$ lead to the detection of a greater number of bursts. Conversely, for smoother and more monotonic distributions, the results are similar across different values of $\alpha$ and $k$.

When analyzing the number of bursts detected in the $65,928$ comment sections with at least 50 unique labeled users (to avoid ambiguities in the burst detection procedure for smaller samples), and using parameters $s = 1.7$, $\gamma = 0.9$, $\alpha = 0.05$, and $k = 0.5$, we observe a generally low to moderate correlation between burst counts and the scores of various polarization measures. A significant majority of comment sections ($87\%$) exhibit only a single detected burst, consistent with measures indicating that these sections are predominantly unpolarized and largely unimodal —except for the Bimodality Coefficient, which deviates from this trend. 

\begin{figure}[t]
  \centering
\includegraphics[width=\textwidth,height=\textheight,keepaspectratio]{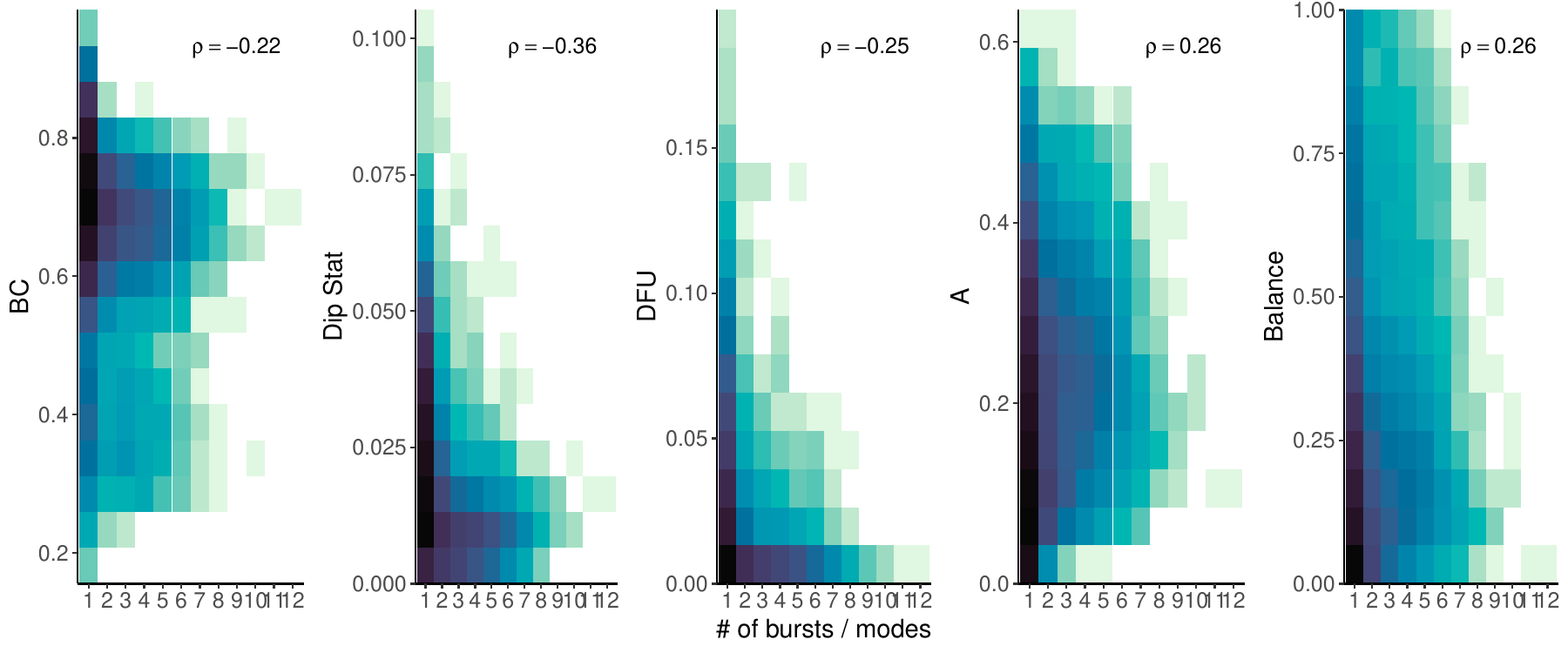}
  \caption{Heatmaps showing the relationship between the number of bursts identified in a comment section, and the score of the different measures. A deeper color indicates a higher volume of data points. $\rho$ refers to the Spearman's Correlation Coefficient between the two variables. While the correlations are generally low to moderate, a higher number of bursts is associated with lower polarization scores—except for the Bimodality Coefficient. This trend is particularly evident for the DFU and Dip Statistic: high polarization scores are typically linked to distributions with two detected modes (bimodal), whereas increasing numbers of detected modes (approaching a uniform distribution) correspond to decreasing polarization scores. This pattern suggests that the burst detection method aligns well with these measures.}
  \label{fig:strip_charts}
\end{figure}

The heatmaps in Figure~\ref{fig:strip_charts} further illustrate the lack of a clear relationship between burst detection outcomes and these measures. Nevertheless, higher burst counts are generally associated with lower polarization scores, particularly for the DFU and Dip Statistic. This observation aligns with the idea that distributions with a higher number of bursts, effectively representing multiple modes, can start to resemble a uniform distribution and are thus considered unpolarized.

A small number of cases exhibit high DFU scores despite only one detected burst. Manual inspection reveals that these distributions often have a dominant peak near one extreme of the range, accompanied by minor peaks scattered across the rest of the range. These minor peaks either go undetected by the burst detection algorithm or are detected but fall below the intensity threshold of 3 imposed by our methodology. As a result, such distributions deviate from the unimodality rule as defined by the DFU, leading to higher polarization scores despite only one significant burst being identified. While these anomalous cases are relatively rare, they underscore the importance of tuning algorithm parameters and accounting for unusual distribution characteristics.

In contrast, the patterns observed with the DFU and Dip Statistic are not reflected in the Bimodality Coefficient, which often classifies distributions with a single detected peak as bimodal, further highlighting the Bimodality Coefficient's limitations in this context. Similar, though less pronounced, inconsistencies are observed with \textit{Balance} and Van der Eijk's $A$. The behavior of \textit{Balance} is expected, as it measures the differential count between two groups rather than capturing deviations from unimodality. For Van der Eijk's $A$, the situation appears once again rather heterogeneous, possibly due to its sensitivity to the distance between modes, which may make it less suitable for use alongside the burst detection procedure.

\section*{Discussion and comments}
In an era where online platforms are increasingly central to news consumption and public discourse, examining political polarization is essential, given its potential role in driving both online and offline conflict. Political polarization is often studied by assessing whether ideological distributions within a population sample depart from a unimodal one. However, despite the widespread use of this framework, there is no consensus on the optimal methodology to measure polarization in this way. 

In this study, we review five commonly used measures, evaluating their strengths and limitations. Our findings indicate that no single measure is universally superior; rather, each has specific advantages depending on the dataset, research objectives, and practical constraints of the analysis. These findings underscore the inherent complexity of reducing a multifaceted phenomenon such as political polarization into a simple, interpretable measure. Although measures can provide valuable information, their effectiveness depends heavily on selecting the right tool for the task at hand, with careful consideration of underlying assumptions and the specific context of analysis.

Through our empirical analysis of political polarization in YouTube comment sections, we observed that four of the five measures generally classify these comment sections as unpolarized, though with varying degrees of consensus. This finding suggests a pattern of ideological segregation in YouTube discussions, with center-leaning videos reasonably showing higher levels of cross-talk between partisans compared to left and right-leaning ones. The degree of variation in the results of the different measures highlights the potential for a situation to be misrepresented due to their characteristics or shortcomings, underscoring the importance of carefully selecting and interpreting them.

Trying to address the limitations of current measures in offering insights regarding the shape of a distribution (especially regarding the number of modes in a distribution), we propose a novel approach that adapts Kleinberg’s burst detection algorithm. When used in combination with - especially - the DFU and Dip Statistic measures, this method enhances their interpretability, offering a more nuanced understanding of ideological structure within online discourse.

Our work represents a step toward a more comprehensive approach to studying political polarization. By systematically evaluating different measures and proposing new methodologies, we aim to contribute to the development of more adaptable and informative tools for analyzing political polarization in online settings.

\section*{Methods}
\subsection*{Generating synthetic samples}

    \paragraph{\textbf{Panel B}} The data utilized in Panel B of Fig. \ref{fig:polarcomp} is synthesized by applying an exponential function with a decay rate of $2.5$ to a linearly spaced range between $−1.0$ and $0$, followed by normalization to ensure the values fall in the range $[-1, 0]$. This results in a smooth, unimodal distribution concentrated near the leftward boundary of the domain.

    \paragraph{\textbf{Panels C, D, E, F}} To generate four of the six synthetic data sets on which the measures are tested in Fig. \ref{fig:polarcomp}, we first randomly generate two independent Gaussian samples with given parameters $\mu_1$, $\sigma_1$, $\mu_2$, $\sigma_2$. These samples are then clipped to lie within the interval $[-1, +1]$. Notice how the clipping procedure alters the distributions, making them no longer strictly Gaussian.
Next, a Bernoulli random variable is utilized as a selection flag, with success probability $p_X(x) \in [0, 1]$, to determine whether a given data point is drawn from the first or the second clipped Gaussian sample. The resulting data set is formed by combining the selected points from both samples. This selection process produces a distribution that resembles a mixture model, where the overall distribution is a probabilistically weighted combination of the two clipped Gaussian distributions, with weights determined by $p_X(x)$.

Regarding Panels C, E and F of Fig.\ref{fig:polarcomp}, the parameters are, respectively:
\begin{itemize}
    \item Panel C: $\mu_1 = -0.5 , \mu_2 = 0.5, \sigma_1 = 1 , \sigma_2 = 1, p_X(x) = 0.25$
    \item Panel E: $\mu_1 = -0.5 , \mu_2 = 0.5, \sigma_1 = 1 , \sigma_2 = 1, p_X(x) = 0.5$
    \item Panel F: $\mu_1 = -0.75 , \mu_2 = 0.75, \sigma_1 = 1 , \sigma_2 = 1, p_X(x) = 0.5$
\end{itemize} 
The sample in Panel D is simply obtained by changing the sign of the one used in Panel C.

\subsection*{Data Collection}
The original list of channels that were considered when selecting videos and scraping the comments was taken from Wu and Resnick (2021) \cite{wu2021cross}, where the authors scraped the bias-checking website ``Media Bias/Fact Check" (MBFC), a commonly used resource in these kind of studies \cite{bovet2019influence, stefanov2020predicting}, to obtain $929$ US-based YouTube political channels for which the leaning was reported. The MBFC dataset was originally supplemented by crawling the ``featured channels" listed on the previously retrieved channels, and by adding three existing YouTube media bias data sets (Lewis (2018) \cite{lewis2018alternative}; Ribeiro \emph{et al.} (2020) \cite{ribeiro2020auditing}; Ledwich and Zaitsev (2020) \cite{ledwich2019algorithmic}).\\
Starting from the original list, by using the official YouTube Data API, we retrieved the ID of every video uploaded by each channel, and filtered them to only keep those uploaded between the 1st of January 2020 and the 31st of August 2020. Subsequently, we filtered out every video which was not tagged as one of these six categories: ``News \& Politics", ``Nonprofits \& Activism", ``Entertainment", ``Comedy", ``People \& Blogs", and ``Education". While seemingly counterintuitive, the categories ``Entertainment" and ``Comedy" were kept as several channels often use these tags in their videos even when they are related to political content. After this, a custom recursive function that leverages the API's functions at our disposal was utilized, allowing us to collect the entirety of the comments posted under each video.\\
Accounting for videos that had their comment section turned off, or that didn't have any comments, we managed to retrieve a total of $76$ million comments posted by $8.8$ million unique users on $159,416$ videos, which were uploaded by $988$ different channels.

\clearpage
\bibliography{bibliography_polars}

\end{document}